\documentclass[12pt]{article}

\usepackage{cite}

\usepackage{amssymb,latexsym}

\usepackage{epstopdf}

\usepackage{graphics,epsfig}

\usepackage{color}

\let\a=\alpha \let\b=\beta \let\g=\gamma \let\d=\delta \let\e=\epsilon
  \let\th=\theta  \let\k=\kappa
\let\l=\lambda \let\m=\mu \let\n=\nu \let\x=\xi \let\p=\pi 
\let\s=\sigma   \let\f=\phi  \let\y=\psi
 
      \let\G=\Gamma  \let\Th=\Theta \let\L=\Lambda
\let\X=\Xi  \let\S=\Sigma  \let\Y=\Psi
 
\let\la=\label  
  
 \def\bd{\begin{document}} \def\ed{\end{document}}
\def\ds{\documentstyle} \let\fr=\frac \let\bl=\bigl \let\br=\bigr
\let\Br=\Bigr \let\Bl=\Bigl
\let\bm=\bibitem
\let\na=\nabla
\def\tU{{\widetilde U}}
\let\pa=\partial \let\ov=\overline
\def\ie{{\it i.e.\ }}
\newcommand{\be}{\begin{equation}}
\newcommand{\ee}{\end{equation}}
\def\ba{\begin{array}}
\def\ea{\end{array}}
\def\ft#1#2{{\textstyle{{\scriptstyle #1}\over {\scriptstyle #2}}}}
\def\fft#1#2{{#1 \over #2}}
\def\F#1#2{{ F_{#1}^{(#2)} }}

\def\R{{\bf R}}
\def\sst#1{{\scriptscriptstyle #1}}
\def\oneone{\rlap 1\mkern4mu{\rm l}}
\def\e7{E_{7(+7)}}
\def\td{\tilde}
\def\wtd{\widetilde}
\def\im{{\rm i}}
\def\bog{Bogomol'nyi\ }
\newcommand{\ho}[1]{$\, ^{#1}$}
\newcommand{\hoch}[1]{$\, ^{#1}$}
\newcommand{\bea}{\begin{eqnarray}}
\newcommand{\eea}{\end{eqnarray}}
\newcommand{\ra}{\rightarrow}
\newcommand{\lra}{\longrightarrow}
\newcommand{\Lra}{\Leftrightarrow}
\newcommand{\ap}{\alpha^\prime}
\newcommand{\bp}{\tilde \beta^\prime}
\newcommand{\cB}{{\cal B}}
\newcommand{\cO}{{\cal O}}
\newcommand{\vecx}{\vec{x}}
\newcommand{\vecy}{\vec{y}}
\newcommand{\vecp}{\vec{p}}
\newcommand{\vecq}{\vec{q}}
\newcommand{\tr}{{\rm tr} }
\newcommand{\Tr}{{\rm Tr} }
\newcommand{\NP}{Nucl. Phys. }

\newcommand{\cL}{{\cal L}}
\newcommand{\cA}{{\cal A}}
\newcommand{\cT}{{\cal T}}
\newcommand{\cF}{{\cal F}}
\newcommand{\cD}{{\cal D}}
\newcommand{\cH}{{\cal H}}
\newcommand{\cM}{{\cal M}}

\def\sst#1{{\scriptscriptstyle #1}}
\def\0{{\sst{(0)}}}
\def\1{{\sst{(1)}}}
\def\2{{\sst{(2)}}}
\def\3{{\sst{(3)}}}
\def\4{{\sst{(4)}}}
\def\5{{\sst{(5)}}}
\def\6{{\sst{(6)}}}
\def\7{{\sst{(7)}}}
\def\8{{\sst{(8)}}}
\def\9{{\sst{(9)}}}
\def\p{{\sst{(p)}}}
\def\q{{\sst{(q)}}}
\def\ve{\varepsilon}
\def\vf{\varphi}
\def\F{\Phi}
\def\wg{\wedge}

\def\thb{\bar{\theta}}
\def\Thb{\bar{\Theta}}
\def\barp{\bar{p}}
\def\barq{\bar{q}}
\def\barc{\bar{c}}
\def\bard{\bar{d}}
\def\e{\epsilon}

\def \bi{\bibitem}
\def \la {\label}

\def \l {\lambda}
\def\foot{\footnote}
\def \tl  {{\tilde \l}}
\def \sql {{\sqrt \l}}
\def \adss {$AdS_5 \times S^5$\ }
\newcommand{\rf}[1]{(\ref{#1})}
\def \ov {\over}

\def\th{\theta}
\def\Th{\Theta}
\def\vth{\vartheta}
\def\btheta{{\bar\theta}}
\def\ttheta{{{\tilde\theta}}}
\def\bttheta{{{\bar\ttheta}}}
\def\vth{\vartheta}

\def\ra{\rightarrow}
\def\N{\nabla}
\def\F{{\cal F}}
\def\uM{\underline{M}}
\def\uA{\underline{A}}
\def\uN{\underline{N}}
\def\uP{\underline{P}}
\def\ua{\underline{a}}
\def\ub{\underline{b}}
\def\uc{\underline{c}}
\def\ud{\underline{d}}
\def\ue{\underline{e}}
\def\uf{\underline{f}}
\def\ui{\underline{i}}
\def\uj{\underline{j}}
\def\uk{\underline{k}}
\def\ul{\underline{l}}
\def\ual{\underline{\alpha}}
\def\ube{\underline{\beta}}
\def\um{\underline{m}}
\def\un{\underline{n}}
\def\up{\underline{p}}
\def\uq{\underline{q}}
\def\ur{\underline{r}}
\def\us{\underline{s}}
\def\umu{\underline{\mu}}
\def\unu{\underline{\nu}}
\def\ula{\underline{\l}}
\def\uka{\underline{\k}}
\def\usi{\underline{\s}}
\def\urh{\underline{\r}}
\def\cc{\circ}
\def\eqv{\equiv}

\def\ni{\noindent}

\def\Ep{E^{{}^{(+)}}}
\def\Em{E^{{}^{(-)}}}

\def\Mp{M^{{}^{(+)}}}
\def\Mm{M^{{}^{(-)}}}

\def \ha{{1\ov 2}}

\def\r{\rho}

\def\Y{{\rm Y}}
\def\X{{\rm X}}
\def\tY{\tilde{\rm Y}}
\def\tX{\tilde{\rm X}}
\def\dY{\dot{\rm Y}}
\def\dX{\dot{\rm X}}

\def \J {\mathcal{J}}
\def \del {\partial}

\def\dF{\dot{F}}
\def\dG{\dot{G}}
\def\df{\dot{f}}
\def \E {{\cal E}}
\def \S {{\cal S}}
\def \J {{\cal J}}

\def\ms{\mathcal{S}}
\def\mj{\mathcal{J}}
\def\soj{\fr{\ms}{\mj}}
\def \R {{\bf R}}
\def \om {\omega}
\def \bE {\bar E}
\def \x {{\cal X}}

\def \bi{\bibitem}
\def \la {\label}

\def \l {\lambda}
\def\foot{\footnote}
\def \tl  {{\tilde \l}}
\def \sql {{\sqrt \l}}
\def \adss {$AdS_5 \times S^5$\ }
\def \ov {\over}

\def \varpi {{\rm w}}

\def\thb{\bar{\theta}}
\def\Thb{\bar{\Theta}}
\def\mb{\bar{\m}}
\def\ab{\bar{\a}}
\def\zb{\bar{z}}
\def\psib{\bar{\psi}}
\def\barp{\bar{p}}
\def\barq{\bar{q}}
\def\barc{\bar{c}}
\def\bard{\bar{d}}
\def\e{\epsilon}
\def\wb{\bar{w}}
\def\lb{\bar{\l}}
\def\Jb{\bar{J}}
\def\Nb{\bar{N}}
\def\Zb{\bar{Z}}
\def\pab{\bar{\pa}}

\def\At{\tilde{A}}
\def\Bt{\tilde{B}}
\def\Ct{\tilde{C}}
\def\Dt{\tilde{D}}
\def\Et{\tilde{E}}
\def\Ft{\tilde{F}}
\def\Gt{\tilde{G}}
\def\Ht{\tilde{H}}
\def\Mt{\tilde{M}}
\def\Rt{\tilde{R}}
\def\at{\tilde{a}}
\def\bt{\tilde{b}}
\def\ct{\tilde{c}}
\def\dt{\tilde{d}}
\def\et{\tilde{e}}
\def\ft{\tilde{f}}
\def\htil{\tilde{h}}
\def\gt{\tilde{g}}
\def\mt{\tilde{\mu}}
\def\nt{\tilde{\nu}}
\def\pht{\tilde{\f}}

\def\rht{\tilde{\rho}}

\def\asth{\hat{*}}
\def\phh{\hat{\phi}}

\def\bA{{\bf A}}

\def\ola{\overleftarrow}
\def\ora{\overrightarrow}
\def\alt{\tilde{\a}}

\def\eh{\hat{e}}
\def\eph{\hat{\e}}
\def\ph{\hat{p}}
\def\alh{\hat{\a}}
\def\beh{\hat{\b}}
\def\gah{\hat{\g}}
\def\Fh{\hat{F}}
\def\muh{\hat{\m}}
\def\nuh{\hat{\n}}
\def\thh{\hat{\th}}
\def\rhh{\hat{\r}}
\def\dh{\hat{d}}
\def\ih{\hat{i}}
\def\jh{\hat{j}}
\def\kh{\hat{k}}
\def\hh{\hat{h}}
\def\nh{\hat{n}}
\def\deh{\hat{\d}}
\def\wh{\hat{w}}
\def\lah{\hat{\l}}
\def\Ah{\hat{A}}
\def\Ch{\hat{C}}
\def\Omh{\hat{\Omega}}

\def\xh{\hat{x}}

\def\ps{\rlap{\, /}\;\,p }
\def\ks{\rlap{\, /}\;\,k }

\def\gym{g_{YM}}

\def\adot{\dot{a}}
\def\bdot{\dot{b}}
\def\bpa{\bar{\pa}}

\def\pr{\prime}
\def\ssk{\medskip}
\def\clb{\color{blue}}
\def\clr{\color{red}}
\def\clv{\colo{violet}}

\def\cV{\mathcal{V}}
\def\cU{\mathcal{U}}

\def\k{\kappa}\def\l{\lambda}\def\L{\Lambda}\def\s{\sigma}\def\S{\Sigma}
\def\Th{\Theta}\def\th{\theta}\def\om{\omega}\def\Om{\Omega}\def\G{\Gamma}
\def\y{\vartheta}\def\m{\mu}\def\n{\nu}
\def\ws{worldsheet}
\def\susy{supersymmetry}
\def\ts{target superspace}
\def\ks{$\k$--symmetry}
\renewcommand{\thefootnote}{\arabic{footnote}}

\renewcommand\baselinestretch{1.2}

\begin{document}

\thispagestyle{empty}

\begin{center}
{\Large \bf
Unifying the PST and the auxiliary tensor \\
\vspace{0.2cm}
field formulations of $4D$ self-duality}

\vspace{1.5cm}

{\bf E.A. Ivanov$^{\flat}$, A.J.
Nurmagambetov$^{\natural}$, and B.M. Zupnik$^{\flat}$}
\vspace{1.5cm} \\ {\it
$^{\flat}$  Bogoliubov Laboratory of Theoretical Physics, JINR},
\\ {\it Dubna, Moscow Region, 141980, Russia}
\\ {\tt eivanov@theor.jinr.ru ~~~~~ zupnik@theor.jinr.ru}
\vspace{0.3cm}\\  {\it
$^{\natural}$ Akhiezer Institute for Theoretical Physics of NSC KIPT},
\\ {\it Kharkov, UA 61108, Ukraine }
\\{\tt ajn@kipt.kharkov.ua}

\end{center}

\medskip
\medskip
\bigskip

\bigskip

\begin{abstract}
\noindent We unify the Lorentz- and $O(2)$ duality-covariant approach to $4D$ self-dual theories by Pasti, Sorokin
and Tonin (PST) with the formulation involving an auxiliary tensor field. We present the basic features
of the new hybrid approach, including symmetries of the relevant generalized PST action. Its salient peculiarity
is the unique form of the realization of the PST gauge symmetries. The corresponding transformations
do not affect the auxiliary tensor field, which guarantees the self-duality of the nonlinear actions in which the $O(2)$
invariant interactions are constructed out of the tensor field.
\end{abstract}

\bigskip
\bigskip
\begin{flushright}
\date{\today}
\end{flushright}


\setcounter{page}{0}
\newpage

\ssk
\ssk
\section{Introduction}

Self-duality is one of the central concepts of gauge theories and string theory. The notorious examples of self-dual $4D$
systems are the renowned Born-Infeld theory and other duality-invariant models of nonlinear electrodynamics.
Reconciling the manifest symmetry under duality rotations \cite{Deser:1976iy}, \cite{Schwarz:1993vs},
\cite{HT:1988} with the manifest Lorentz invariance becomes possible in the formulations
with auxiliary fields \cite{Pasti:1995ii}, \cite{Pasti:1995tn},
\cite{Bengtsson:1996fm}, \cite{Berkovits:1996rt}, \cite{Pasti:1996vs}.
The most economic approach
requires just a single scalar auxiliary field entering the action
non-polynomially.
This formulation was originally developed for the free self-dual tensor fields
\cite{Pasti:1995ii}, \cite{Pasti:1995tn}, \cite{Pasti:1996vs} and, later on,
was extended to nonlinear models
of branes and their coupling to supergravity actions
\cite{Pasti:1997gx}, \cite{Bandos:1997ui}, \cite{Aganagic:1997zq},
\cite{Bandos:1997gd}, \cite{Nurmagambetov:1998gp}.
Introducing the interaction into the self-dual covariant
actions is a non-trivial task as the interaction terms should satisfy a
consistency condition generalizing that of \cite{Gaillard:1981rj}.

\ssk
\ssk
On the other hand, there exists a universal approach to duality-invariant
$4D$ theories based on employing the auxiliary
tensor (bispinor) fields \cite{Ivanov:2003uj},
\cite{Ivanov:2012bq}, \cite{Ivanov:2013ska}.
It came out as a by-product of studying ${\cal N}=3, 4D$ Born-Infeld theory in the harmonic
superspace formulation \cite{Ivanov:2001ec}. Within this approach,
the interaction part of the action is constructed solely out of the auxiliary tensor fields and
is manifestly duality-invariant. Though the whole action is not duality-invariant,
on shell it leads to the equations of motion which, together with the Bianchi identity, are covariant under
the duality rotations. After elimination of the auxiliary fields by their equations of motion,
the resulting system automatically obeys the general nonlinear Gaillard-Zumino
consistency condition \cite{Gaillard:1981rj}. The whole set of the self-dual actions
of nonlinear $4D$ electrodynamics thus proves to be in the one-to-one correspondence with the appropriate auxiliary interactions.
In refs. \cite{Ivanov:2003uj}, \cite{Ivanov:2012bq}, \cite{Ivanov:2013ska} various nonlinear self-dual
models were explicitly constructed in this way. The supersymmetric versions of the approach with the auxiliary tensor fields
were worked out in \cite{N1N2}.

\ssk
\ssk
The aim of the present Letter is to elaborate on a new formulation of the self-dual $4D$ actions with tensor fields,
such that they enjoy the manifest duality invariance off shell.
This goal is pursued by properly extending the construction
of Pasti-Sorokin-Tonin (PST) \cite{Pasti:1995ii}, \cite{Pasti:1995tn},
\cite{Pasti:1996vs}. The striking feature of the hybrid formulation is that the renowned
PST gauge symmetry transformations have a universal form, irrespective of the precise structure of the self-interaction.

\ssk
\ssk
The Letter is organized as follows. In Section 2 we recall the structure
and the symmetries of the original PST action of the free duality-symmetric Maxwell field in $4D$.
In Section 3 we extend the PST formulation by introducing an auxiliary tensor field
and discuss the duality invariance, as well as the symmetry structure of the proposed action. It allows a direct generalization
to the interaction case, without affecting the form of the PST gauge transformations.
Section 4 contains a brief discussion of the relations of the new action to the previously known
non-covariant duality-symmetric actions of the $4D$ Maxwell field.
Summary and conclusions are collected in the final part of the paper.

\ssk
\ssk
\section{PST action and its symmetries}

We start with a brief discussion of the standard PST approach to 4D
self-dual theories within the original second order formulation of
\cite{Pasti:1995ii}, \cite{Pasti:1995tn}, \cite{Pasti:1996vs}.
The following action\footnote{We use the  conventions
$g_{mn}=\mbox{diag}(1,-1,-1,-1),\quad \epsilon_{0123}=1$.}
\be
S_{PST}=\fr 12 \int \,d^4 x \, \left[- v^m \Ft^{a}_{mn}\,\d_{ab} \,
\Ft^{b\, nl}\,v_{l}+ v^m F^{b}_{mn}\,\e_{ab}\, \Ft^{a\,  nl}\,v_l \right]
\la{PSTso}
\ee
describes the dynamics of the duality-symmetric Maxwell field
$A^{a}_m$, $a=1,2$, with the field strength
$F^{a}_{mn}=\pa_mA_{n}^a-\pa_nA_{m}^a$. Its dual is defined by
\be
\Ft^a_{mn}=\fr12 \e_{mnrl}\, F^{a\,rl} \,.
\la{Fdual}
\ee
The other entities entering the action \rf{PSTso} are the PST scalar $a(x)$
in the specific non-polynomial combination $v_m$
\be
v_m=\fr{\pa_m a(x)}{\sqrt{(\pa a)^2}} 
\la{vdef}
\ee
and the $O(2)$ invariant tensors $\d_{ab}$ and $\e_{ab}$ ($\e_{12}=1$,
$\e_{21}=-1$). The role of the PST scalar is to make the Lorentz covariance
of the action manifest; the $O(2)$ tensors ensure the manifest
invariance of the action under the $O(2)$ duality rotations of the vector
fields $A^{a}_m\,$, $\delta A^{a}_m = \omega \e_{ab}A^{b}_m\,$.

\ssk
\ssk
The general variation of the action \rf{PSTso} is calculated to be
\[
\d S_{PST}=-\fr12 \int\,d^4 x\, \left[\d F^a_{mn}- \d v_m \,
(v\cdot \cF^a)_n \right] \e^{mnrl}\, v_r\,\e_{ab} \,(v\cdot \cF^b)_l
\]
\be
=-\int\,d^4x\, \left(\d A^a_m-\fr{\d a}{\sqrt{(\pa a)^2}}\, (v \cdot \cF^a)_m
\right)\e^{mnrl} \pa_n \left(v_r\, \e_{ab}\, (v\cdot \cF^b)_l \right),
\la{PSTvarg}
\ee
where we have omitted the total derivative term and introduced
\be
\cF^a_{mn}=F^a_{mn}+\e_{ab}\,\Ft^b_{mn}\,,
\la{cFdef}
\ee
with $(v\cdot \cF^a)_n := v^m \,\cF^a_{mn}$. As follows from
\rf{PSTvarg}, the action \rf{PSTso}, besides the invariance under $U(1)$
local gauge transformations of the Maxwell fields $A^a_m$, reveals the invariance
under the following special gauge symmetries (the so-called PST symmetries
\cite{Pasti:1995ii}, \cite{Pasti:1995tn}, \cite{Pasti:1996vs}):
\be
\d_I \,A^a_m=\pa_m a(x)\, \Phi_a(x),\qquad \d_I \,a(x)=0 \,,
\la{PSTsymI}
\ee
\be
\d_{II} \,a(x)=\vf(x),\qquad \d_{II} \,A^a_m=
\fr{\vf(x)}{\sqrt{(\pa a)^2}}\, (v \cdot \cF^a)_m \,.
\la{PSTsymII}
\ee

\ssk
\ssk
These two PST symmetries (below we refer to them as PST-I and PST-II)
play different roles. The PST-I symmetry is needed to reduce, by fixing the gauge
parameters $\Phi_a(x)$, the vector field equation of motion
\be
\e^{mnrl} \pa_n \left[v_r\, \e_{ab}\, (v\cdot \cF^b)_l \right]=0
\la{Aeom}
\ee
to the self-duality condition\footnote{Details of this procedure may be found, e.g., in
\cite{Sorokin:2002}, or, more recently, in \cite{Pasti:2012wv}.}
\be
\cF^a_{mn}=F^a_{mn}+\e_{ab}\,\Ft^b_{mn}=0\,.
\la{sd}
\ee

The second PST symmetry (PST-II) guarantees that the presence of the PST scalar
$a(x)$, which is needed for the manifest Lorentz covariance of the PST action, does not
increase the number of the initial degrees of freedom: This field can be completely gauged away. Indeed, its equation of motion
\be
\pa_m \Big[\fr{1}{\sqrt{(\pa a)^2}}\, \e^{mnrl}\,v_n\, \e_{ab}\,
(v\cdot \cF^a)_r \,(v\cdot \cF^b)_l \Big]=0
\la{aeom}
\ee
does not contain any additional information and is trivially satisfied on shell.
However, a gauge fixing of the PST scalar breaks the manifest Lorentz covariance
of the model, resulting in the manifestly duality-invariant but non-covariant
formulation of Schwarz and Sen \cite{Schwarz:1993vs} (see Sect. 4 for details).

\ssk
\ssk
\section{Manifestly covariant self-dual action with tensor auxiliary fields}

As shown in \cite{Ivanov:2003uj}, \cite{Ivanov:2012bq},
\cite{Ivanov:2013ska}, introducing the auxiliary
tensor fields allows one to drastically simplify the problem of finding the $O(2)$ duality-invariant
interactions. It was of obvious interest to generalize the auxiliary tensor field formulation in such a way
that the $O(2)$ duality becomes the manifest off-shell symmetry of the total action.

\ssk
This goal motivated us to consider the following modification of the PST action \rf{PSTso}
\[
S= \int\, d^4 x\,  \cL_{PST}' \equiv \int\, d^4x\, \,\left[ \cL_{PST}+v^m \Ft^a_{mn}\,\d_{ab} \,
\Ft^{b\, nl}\,v_{l} \right.
\]
\be
\left. + v^m\,{V}_{mn}\,{V}^{nr}v_r
+v^m\,\tilde{V}_{mn}\, \tilde{V}^{nr}v_r
+2v^m\, V_{mn} \,\Ft^{2nl}\,v_l-2v^m\, \tilde{V}_{mn}\,\Ft^{1nl}\,v_l \right].
\la{PSTfo}
\ee
Integrating out an unconstrained auxiliary field $V_{mn}$ takes as back to
the original PST action. Indeed, the equations of motion for $V_{mn}$ read
\be
v_{[m}\left(V_{n]p} + \tilde{F}^2_{n]p}\right)v^p
-\frac12 v^t\e_{tsmn}\left(\tilde{V}^{sp}
- \tilde{F}^{1\,sp} \right)v_p = 0\,,
\ee
whence
\be
v^mV_{mn} = -v^m\tilde{F}^2_{mn}\,, \quad v^m\tilde{V}_{mn} =
v^m\tilde{F}^1_{mn}\,.\la{solut}
\ee
Substitution of these expressions into \rf{PSTfo} leaves us with
${\cal L}_{PST}$ as the ``on-shell'' Lagrangian.
Note that the relations \rf{solut} in fact enable to express the whole
$V_{mn}$ in terms of $v_m$ and the field strengths $F^a_{mn}$ (the number of
the independent relations in \rf{solut} is just 6, and $V_{mn}$ has 6
independent components). It is convenient to present
the corresponding expressions in the bispinor notation, using the definitions
\bea
&& V_{mn} -i\tilde{V}_{mn} = (\tilde\sigma_m\sigma_n)_{\dot\beta}^{\dot\alpha}\bar{V}^{\dot\beta}_{\dot\alpha}\,, \quad V_{mn} +i\tilde{V}_{mn} =
-(\sigma_m\tilde\sigma_n)^\alpha_\beta V_\alpha^\beta\,,  \nonumber \\
&& v_m = \frac12(\tilde{\sigma}_m)^{\dot\alpha\beta}v_{\beta\dot\alpha}
=\frac12({\sigma}_m)_{\beta\dot\alpha}\tilde{v}^{\dot\alpha\beta}\,,
\quad v_{\beta\dot\alpha}\tilde{v}^{\dot\alpha\rho}=\delta^\rho_\beta\,,
\eea
and the similar ones for $F^a_{mn}$. Then eqs. \rf{solut} amount to the set
\bea
&& v_{\beta\dot\xi}V_\xi^\beta +v_{\xi\dot\alpha}\bar{V}^{\dot\alpha}_{\dot\xi}
=i[v_{\beta\dot\xi}(F^2)_\xi^\beta -v_{\xi\dot\alpha}(\bar{F}^2)^{\dot\alpha}_{\dot\xi}]\,, \nonumber \\
&& v_{\beta\dot\xi}V_\xi^\beta
-v_{\xi\dot\alpha}\bar{V}^{\dot\alpha}_{\dot\xi}
=v_{\beta\dot\xi}(F^1)_\xi^\beta
-v_{\xi\dot\alpha}(\bar{F}^1)^{\dot\alpha}_{\dot\xi}\,,
\eea
and hence we find
\bea
V^{\alpha}_{\xi} =: I^{\alpha}_{\xi}(F)=\frac12(F^1-iF^2)^{\alpha}_{\xi}
-\frac12v_{\xi\dot\beta}\tilde{v}^{\dot\rho\alpha}
(\bar{F}^1-i\bar{F}^2)_{\dot\rho}^{\dot\beta}\,,\quad
\bar{V}^{\dot\alpha}_{\dot\xi} = \overline{(V^{\alpha}_{\xi})}\,.
\label{bispsol}
\eea
This solution could be equivalently derived in a more direct way, starting from the action
\rf{PSTfo} in the bispinor notation.

\ssk

The extended action \rf{PSTfo}, like its pure PST prototype, is $O(2)$ duality invariant. Indeed, the involved quantities possess
the following $O(2)$ transformation properties:
\be
\delta F^a_{mn}=\omega \e_{ab}F^b_{mn}\,, \quad  \delta V_{mn}=\omega\tilde{V}_{mn} \Leftrightarrow \delta \tilde{V}_{mn} = -\omega V_{mn}\,.
\ee
One can join $V_{mn}, \tilde{V}_{mn}$ into the doublet $V^a_{mn} := (V_{mn}, \tilde{V}_{mn}), \;\tilde{V}^a_{mn} = \e_{ab} V^b_{mn}\,,$
and rewrite the $V$-dependent terms in \rf{PSTfo} in the manifestly $O(2)$ invariant form
\be
v^m V^a_{mn}\d_{ab} V^{b nr}v_r + 2 v^m V_{mn}^a\e_{ab} \tilde{F}^{b nl}v_l\,.
\ee
We observe that for preserving the manifest duality invariance it is enough to add a {\it single} auxiliary tensor field,
still keeping a double set of the gauge potentials.

\ssk
Thus by construction the action \rf{PSTfo} is manifestly Lorentz- and
duality-invariant. However, like in the original PST approach
\cite{Pasti:1995ii}, \cite{Pasti:1995tn}, \cite{Pasti:1996vs}, the
covariance of the action is ensured by the PST scalar $a(x)$, the auxiliary nature of which
is guaranteed by the PST symmetries. Therefore,
we are led  to find how (if any) the transformation laws \rf{PSTsymI}, \rf{PSTsymII}
are modified upon introducing the auxiliary tensor field $V_{mn}$.

\ssk
The straightforward computations lead to the following expression for the general variation of \rf{PSTfo}
(modulo a total derivative):
\[
\d S=-\int\,d^4x\, \left(\d A^a_m-\fr{\d a}{\sqrt{(\pa a)^2}}\,
(v \cdot \widehat{\cF}^a)_m \right)\e^{mnrl} \pa_n
\left(v_r\, \e_{ab}\, (v\cdot \widehat{\cF}^b)_l \right)
\]
\be
+2\int\,d^4x\, \d V_{mn} \left(v^m \left[-V^{nr}-\Ft^{2\,nr}\right]v_r
+\fr12 v_s \,\e^{su\,mn} \left[\tilde{V}_{ur}-\Ft^1_{ur} \right]v^r \right),
\la{PSTfogv}
\ee
where
\be
\widehat{\cF}^1_{mn} :=F^1_{mn}- \Ft^2_{mn}-2V_{mn}\,,
\qquad \widehat{\cF}^2_{mn}=F^2_{mn}+ \Ft^1_{mn}-2 \tilde{V}_{mn}\,.
\la{whFF'def}
\ee
The variation \rf{PSTfogv} vanishes under the standard local $U(1)$ transformations of the gauge fields $A^a_m$,
as well as under the following modified PST-type transformations
\be
\d_I \,A^a_m=\pa_m a(x)\, \Phi_a(x),\qquad \d_I \,a(x)=0 \,,
\qquad \d_I\, V_{mn}=0\,,
\la{PSTsymIg}
\ee
\be
\d_{II} \,a(x)=\vf(x),\qquad \d_{II} \,A^a_m=
\fr{\vf(x)}{\sqrt{(\pa a)^2}}\, (v \cdot \widehat{\cF}^a)_m\,,
\qquad  \d_{II}\, V_{mn}=0\,.
\la{PSTsymIIg}
\ee

\ssk
We see that only the PST-II transformations are actually modified. It is very important to note
that the modified PST transformations
\rf{PSTsymIg}, \rf{PSTsymIIg} do not affect the auxiliary tensor field
$V_{mn}$. This peculiarity has a great  impact on the structure of admissible
interaction terms.

Just due to this notable property, in the nonlinear case we can add, to the bilinear action \rf{PSTfo},
an arbitrary  auxiliary interaction ${\cal E}(A)=\frac12 A+O(A^2)\,$, where $A$ is the quartic $O(2)$ invariant
variable
\be
 A=(\mbox{Tr}V^2)(\mbox{Tr}\bar{V}^2)
=\frac1{16}[(V^{mn}V_{mn})^2+(V^{mn}\tilde{V}_{mn})^2], \la{Variable}
\ee
with $(\mbox{Tr}V^2)=V^\beta_\alpha V_\beta^\alpha=
\overline{(\mbox{Tr}\bar{V}^2)}$.
The free solution $V_\beta^\alpha=I_\beta^\alpha (F)$ (\ref{bispsol}) can be easily generalized to
the interaction case
\bea
V^\beta_\alpha=I^\beta_\alpha(F)-(v\bar{V}\tilde{v})^\beta_\alpha(\mbox{Tr}V^2)
{\cal E}_A\,, \la{AuxEq}
\eea
where ${\cal E}_A=d{\cal E}/dA\,$. The resulting action
\bea
S_{int} = \int\,d^4x\,[{\cal L}_{PST}' + {\cal E}(A)] \la{interCov}
\eea
preserves all symmetries of the bilinear action,
including the gauge PST symmetries. Thus it describes some self-dual system for any choice of the $O(2)$ invariant interaction.
 After solving eq. \rf{AuxEq} (e.g., by recursions), one is left with the highly nonlinear action in terms of the field
 strengths $F^a_{mn}$ and the auxiliary scalar field $a(x)\,$\footnote{The $O(2)$ invariant
interactions with derivatives of $V_{mn}$ are also admissible, leading to self-dual actions with derivatives
on the gauge field strengths \cite{Ivanov:2013ska}.}. The PST-II transformations also become nonlinear.  It is remarkable
 that, before eliminating $V_{mn}$, the PST transformations have the universal form \rf{PSTsymIg},  \rf{PSTsymIIg}.

\ssk
\ssk
\section{Non-covariant self-dual action with tensor auxiliary field}

Let us briefly discuss the relation of the covariant action \rf{PSTfo} to
the previously proposed duality-invariant actions of \cite{Schwarz:1993vs},
\cite{Rocek:1997hi}, \cite{Ivanov:2012bq}.

\ssk
The PST-II symmetry of the action \rf{PSTsymIIg} makes it possible to fix
the gauge $v_m=\d^0_m$. As a result, the covariant action \rf{PSTfo} turns
into the following duality-symmetric non-covariant action:
\be
S_{n.c.}=\int\,d^4 x\, \left[\fr12 B^a_k \,\d_{ab}\, B^b_k+\fr12 B^a_k \,
\e_{ab}\,E^b_k+2V^a_k\, \e_{ab}\,B^b_k+V^a_k\, \d_{ab}\,V^b_k \right].
\la{LVEBnc}
\ee
Here, we have introduced $V_{0i}=V^1_i=V_i$, ~$\tilde{V}_{0i}=V^2_i=U_i$
\footnote{4D indices are split into the $1+3$ set as $m=(0,i)$.
We denote $F^a_{0i}=E^a_i$,~ $F^a_{ij}=\e_{ijk}B^a_k$,~
$\tilde{F}^a_{0i}=B^a_i$;
in our notation
$\e^{0123}=-1$, so $\e^{0ijk}=-\e_{ijk}$ and $\e_{0ijk}=\e_{ijk}$.}. The action
\rf{LVEBnc} is a non-covariant gauge-fixed ``magnetic'' version of the covariant action
proposed in \cite{Rocek:1997hi} (see Appendix C of \cite{Ivanov:2012bq} for details of
deriving \rf{LVEBnc}\footnote{In fact, in \cite{Ivanov:2012bq} an ``electric'' version of \rf{LVEBnc}
was derived, using the trick suggested in \cite{Rocek:1997hi}; the action \rf{LVEBnc} follows from the ``electric'' version
through a discrete  duality transformation.}).

\ssk
After integrating out the auxiliary fields $V^a_k\,$, the action \rf{LVEBnc} takes the form
\be
S_{SS}=\int\,d^4x\,\left[ -\fr12 B^a_k\, \d_{ab}\, B^b_k+\fr12 B^a_k\,
\e_{ab}\,E^b_k \right],
\la{SSac}
\ee
which is none other than  the Schwarz-Sen non-covariant duality-invariant
action \cite{Schwarz:1993vs}.
This result is of course not surprising, because the elimination of
the tensor auxiliary field in the gauge-unfixed action \rf{PSTfo} takes the latter just back to the PST action. On the other hand,
the PST action \rf{PSTso} is a covariantization of the Schwarz-Sen action
\rf{SSac}.

\ssk
Extending the action \rf{LVEBnc} to the non-trivial interaction comes about as
follows
\be
L_{n.c.}=
\fr12 B^a_k \,\d_{ab}\, B^b_k+\fr12 B^a_k \,
\e_{ab}\,E^b_k+2V^a_k\, \e_{ab}\,B^b_k+V^a_k\, \d_{ab}\,V^b_k +
\mathcal{E}(A)\,,
\la{LVEBncE}
\ee
where $\mathcal{E}(A)$ is the same function of the manifestly $O(2)$ duality-invariant
variable \rf{Variable} as in the covariant action \rf{interCov}. In the $3D$ notation it
is constructed out of the $3D$ components of $V_{mn}$ as
\bea
A=\frac14(U_kU_k)^2+\frac14(V_kV_k)^2-\frac12(U_kU_k)(V_iV_i)+(V_iU_i)^2.
\eea
The corresponding auxiliary equations are
\[
V^1_k+B^{2}_k+\frac12\frac{\partial A}{\partial V^1_k}{\cal E}_A=0,
\qquad {\cal E}_A\equiv \fr{\pa \mathcal{E}(A)}{\pa A}\,,
\]
\be
V^2_k-B^{1}_k+\frac12\frac{\partial A}{\partial V^2_k}{\cal E}_A=0\,.
\la{VsdE}
\ee
The general non-covariant action \rf{LVEBncE} can be obtained as a gauge-fixed version of the general
covariant action \rf{interCov} which enjoys both PST gauge symmetries.

\ssk
\ssk
\section{Conclusions}

To summarize, we have proposed the new approach to the $4D$ self-dual nonlinear electrodynamics systems, which is a symbiosis of
the Pasti-Sorokin-Tonin covariant duality-invariant approach \cite{Pasti:1995ii}, \cite{Pasti:1995tn},
\cite{Pasti:1996vs} and the approach of \cite{Ivanov:2003uj} ,
\cite{Ivanov:2012bq}, \cite{Ivanov:2013ska} involving auxiliary tensor fields. The new approach inherits
the advantages of both approaches just mentioned. On the one hand, it is manifestly Lorentz and $O(2)$ duality
covariant. On the other hand, it provides a simple way of constructing  self-dual actions with a non-trivial interaction.
We have studied the symmetry structure of the proposed action and established its relation
to the duality-symmetric approaches of \cite{Schwarz:1993vs}, \cite{Rocek:1997hi}. The most sound feature of the
action constructed is the universal form of the gauge PST transformations off shell, before eliminating the
auxiliary tensor fields by their equations of motion. These transformations do not affect the auxiliary fields at all,
the feature that makes it possible to construct invariant interactions from these fields without breaking
any symmetry of the free action. Note that PST actions with additional auxiliary fields were considered before
(see, e.g., \cite{Pasti:1996vs}, \cite{Maznytsia:1998xw}), however the approach we follow here is entirely different,
since it is not related to any dualization of the PST scalar field.

\ssk
An important feature of our formulation is that it ensures a consistent way of adding a non-trivial interaction to the free
actions, with the guarantee that the emerging nonlinear system is self-dual. Recently, the general structure of nonlinear
interacting self-dual actions within the PST approach was analyzed in \cite{Pasti:2012wv}. It was found there
that the invariance of the whole action under the PST-type transformations amounts to the fundamental consistency condition
of the Gaillard-Zumino type \cite{Gaillard:1981rj}. It would be interesting to establish the precise links of our
approach with this general analysis.

\ssk

\bigskip
\ssk
\ssk
\noindent{\bf Acknowledgements.} This work was supported in part by the Joint
DFFD-RFBR Grant \# F53.2/012 (Ukraine) and \# 13-02-90430
(Russian Federation). E.I. \& B.Z. acknowledge a support from the RFBR grants Nr.12-02-00517,
Nr.13-02-91330, the grant DFG LE 838/12-1 and a grant of the Heisenberg-Landau program.
The authors thank Dima Sorokin for valuable discussions and correspondence. E.I. thanks Arkady Tseytlin for useful correspondence.
A.N. is grateful to BLTP JINR for the kind hospitality during  the final stage of this study.

\end{document}